\documentclass[aps,twocolumn,prb]{revtex4}

\usepackage{graphicx}

\usepackage{amsmath}
\usepackage{bm}

\begin{document}

\title{Tailoring the excitation of localized surface plasmon-polariton 
resonances\\ by focusing radially-polarized beams}

\author{Nassiredin M.~Mojarad}
\email{nassiredin.mojarad@phys.chem.ethz.ch} 
\homepage{www.nano-optics.ethz.ch} 
\author{Mario Agio}

\affiliation{Nano-Optics Group, Laboratory of Physical Chemistry, ETH 
Zurich\\ CH-8093 Zurich, Switzerland}


\begin{abstract}
We study the interaction of focused radially-polarized light with
metal nanospheres. By expanding the electromagnetic field in terms
of multipoles, we gain insight on the excitation of localized 
surface plasmon-polariton
resonances in the nanoparticle. We show that focused radially-polarized 
beams offer more opportunities than a focused plane wave
or a Gaussian beam for tuning the near- and far-field system response.
These results find applications in nano-optics, optical tweezers, and
optical data storage.
\end{abstract}


\maketitle

\section{Introduction}
Focused radially-polarized beams (FRBs) find application in many
fields of optics. For instance, it has been demonstrated that they
achieve a tighter focal spot compared to focused plane-wave (FPW)
illumination~\cite{leuchs1,leuchs2}. Moreover, Novotny {\it et
al.} have used them to map the orientation of single
molecules~\cite{novotny}. Van Enk has discussed FRBs in the
context of quantum optics for efficient coupling of light to a
single emitter~\cite{enk}. Also in the area of optical tweezers
Zhan has shown that metal nanoparticles (NPs) can be trapped
in three dimensions under FRBs~\cite{zhan1}. Meixner's group has
performed a set of experiments to explore the
interaction of radially-, azimuthal- and linearly-polarized focused
beams with metal NPs~\cite{meixner1,meixner2,meixner3}.

Recently, a number of theoretical works have studied the
interaction of tightly-focused beams with spherical NPs. \c{S}endur {\it
et al.} have used diffraction integrals to analyze the near field
produced by a silver NP under FPW and FRB
illumination~\cite{sendur}. Likewise, van de Nes and T\"or\"ok
have developed a generalized Mie theory for Gauss-Laguerre beams,
including the case when the NP is displaced from the
focus~\cite{torok2}. More recently, Moore and Alonso have
chosen the multipole expansion approach to develop a generalized
Mie theory where the incident beam is a complex focused field. In
particular, they study the response of a dielectric NP under FRBs
at a fixed wavelength~\cite{italians2}. Lastly, Lerm\'{e} {\it et
al.} have studied the extinction of light by silver NPs in the
context of the spatial modulation spectroscopy
technique~\cite{lerme}.

In this paper we extend our earlier study on the localized 
surface plasmon-polariton (LSP)
spectra of metal NPs~\cite{JOSAB} to the case of FRBs. Using the multipole
expansion approach, we show that, contrary to a FPW~\cite{torok} and a
Gaussian beam~\cite{lock},
FRBs offer more opportunities for manipulating the excitation of higher-order
LSP resonances in metal NPs. We discuss these features both in the
far- and near-field regimes.

\section{Generalized Mie theory}

Radially-polarized doughnut beams are constructed by superposing
the Gauss-Hermite modes with normal polarizations~\cite{oron}. The
effect of the lens is identified by two parameters, the focusing
semi-angle $\alpha$ and the $a$ factor defined as
$f/w_o$~\cite{enk}, where $f$ is the focal length of the lens and
$w_o$ is the beam waist (see inset in Fig.\ref{multipole}). Using
the Richards and Wolf formalism for an aplanatic
system~\cite{richards} and adapting it to radially-polarized
illumination, the field right after passing
the lens is given by $\mathbf{E}(a,\theta)=A
\exp(-a^2\sin^2\theta) f\sin\theta \sqrt{\cos\theta} \hat{\bm
\theta}$, where $\sqrt{\cos\theta}$ represents the apodization
function and $\hat{\bm \theta}$ gives the orientation of the electric field 
in spherical coordinates ($r,\theta,\phi$)~\cite{leuchs1,brown}.
Note that this field depends only on $a$ and $\theta$ because $f$
can be merged into the amplitude $A$, which only rescales the intensity
of the beam.

The electric field in the image space can now be found by the
multipole expansion, where the weight coefficient of each
multipole is determined by matching the field at the lens
boundary~\cite{JOSAB,torok,italians1}. Since the electric
field has only a $\theta$ component in the far field, the
azimuthal number is $m=0$,
and the symmetry around the $z$ axis implies that transverse
electric (TE) multipoles have no contribution. The incident field
and the coefficients are hence given by
\begin{eqnarray}
\label{incident} \mathbf{E}_\mathrm{inc} & = &
\sum_{l=1}^{\infty}B_l \mathbf{N}_{e0l}^{(1)},\\
 B_l & = & \frac{2l+1}{2l(l+1)} \frac{2 k f}{i^l \exp(-i k f)}
\times \\ \nonumber & &
\int_{0}^{\alpha}
 |\mathbf{E}(a,\theta)| \frac{\mathrm{d}
P_l(\cos\theta)}{\mathrm{d}\theta}\sin\theta
 \mathrm{d}\theta,
\end{eqnarray}
where $k$ is the wavevector and $P_l(\cos\theta)$ are Legendre
polynomials. Here the multipoles follow the notation
of~\cite{bohren}. The expansion of Eq.~(\ref{incident}) is a
special case of tightly-focused spirally-polarized
beams~\cite{italians1}.

Now that the incident field is determined in terms of
multipoles, the effect of placing a spherical NP at the origin can
be solved by the generalized Mie theory~\cite{JOSAB}. The scattered 
and internal fields are respectively found to be
\begin{equation} \label{scattint}
\mathbf{E}_\mathrm{s}=-\sum_{l=1}^{\infty}a_l B_l
\mathbf{N}_{e0l}^{(3)}, \hspace{1cm}
\mathbf{E}_\mathrm{i}=\sum_{l=1}^{\infty}d_l B_l
\mathbf{N}_{e0l}^{(1)},
\end{equation}
where $a_l$ and $d_l$ are the Mie coefficients as defined in \cite{bohren}. 
Note that because the incident field does not contain TE multipoles, these
are also absent in the scattered and internal fields. Having the
fields expanded in all space, the total scattered ($W_\mathrm{s}$)
and extinguished ($W_\mathrm{e}$) powers can also be calculated by
integrating the corresponding Poynting vectors in the far field,
giving
\begin{eqnarray}
\label{Ws}
W_\mathrm{s} & = & \frac{\pi}{Z k^2×}\sum_{l=1}^{\infty}|B_l|^2|a_l|^2
\frac{2l(l+1)}{2l+1}, \\
\label{We}
W_\mathrm{e} &  = & \frac{\pi}{Z k^2×}\sum_{l=1}^{\infty}|B_l|^2 
\mathrm{Re} \{a_l\}
\frac{2l(l+1)}{2l+1},
\end{eqnarray}
where $Z$ is the impedance of the background medium. According to
Eqs.~(\ref{Ws}) and (\ref{We}), $W_\mathrm{s}$ and $W_\mathrm{e}$
not only depend on the NP size and material, but also vary
with the multipole strength (the $B_l$ term)~\cite{JOSAB}. This
clearly shows that for a specific LSP resonance, one can
in principle control the scattered and extinguished powers of the
corresponding multipole by adjusting
the illumination beam and the focusing system.

The cross sections associated with $W_\mathrm{s}$ and
$W_\mathrm{e}$ are found by dividing these quantities by the
average intensity incident on the $xy$ section of the
NP~\cite{JOSAB}. This averaging process is necessary because
the field in the focal plane is strongly inhomogeneous in contrast
to plane-wave illumination. However, the cross sections
calculated by this approach lead to unrealistic values that are
one or two orders of magnitude larger than those obtained by FPW
illumination. The reason behind this delusive behavior is that
although the electric field is strong near the focus, the Poynting
vector is small and vanishes on axis (see inset in
Fig.~\ref{ext})~\cite{novotny}. We therefore use another quantity
to study the LSP spectra, namely the scattering ($\mathcal{K}_\mathrm{s}$) 
and the extinction ($\mathcal{K}_\mathrm{e}$) efficiencies, defined by 
dividing $W_\mathrm{s}$ and $W_\mathrm{e}$ by the total incident power 
$P_\mathrm{inc}$~\cite{zumofen}: 
$\mathcal{K}_\mathrm{s}=W_\mathrm{s}/P_\mathrm{inc}$ and
$\mathcal{K}_\mathrm{e}=W_\mathrm{e}/P_\mathrm{inc}$.
The profiles obtained by this method slightly differ from the cross sections,
but still exhibit the resonances of the system.

Many applications of plasmonic materials are based on their
near-field properties~\cite{maier}, whereas cross sections or
efficiencies are far-field quantities. For this reason, we use an
alternative definition called the near-field average intensity
enhancement $K$ for characterizing the system response. This
quantity is simply the total field intensity averaged over the
NP surface, divided by the intensity at the origin in the
absence of the NP~\cite{JOSAB}. By taking advantage of the
orthogonality of the multipoles, $K$ is separated into radial
$K_\mathrm{r}$ and tangential $K_\mathrm{t}$ components,
\begin{eqnarray}
\label{radial}
K_\mathrm{r} & = & \frac{9}{16(kR)^4}\sum_{l}
\left|\frac{B_l}{B_1}\right|^2
\frac{l^2(l+1)^2}{2l+1} \times \\ \nonumber
& & \left[|\psi_l|^2 +|a_l|^2|\chi_l|^2 +2
\mathrm{Re} \{a_l^*\chi_l^*\}\psi_l\right],\\
\label{tangential}
K_\mathrm{t} & = & \frac{9}{16(kR)^2}\sum_{l} \left|\frac{B_l}{B_1}\right|^2
\frac{l(l+1)}{2l+1} \times \\ \nonumber
& & \left[|\psi_l'|^2 +|a_l|^2|\chi_l'|^2 +2
\mathrm{Re} \{a_l^*\chi_l'^*\}\psi_l'\right],
\end{eqnarray}
where $R$ is the NP´s radius. In these equations $\psi_l$ and
$\chi_l$ are respectively the Riccati-Bessel functions of the
first and third kinds calculated at the NP surface, and the
prime indicates the derivative with respect to $kr$~\cite{JOSAB,bohren}.

\section{Results and discussion}

In this section we discuss how FRBs can change the excitation of
LSP resonances in metal NPs and its
consequences on the far- and near-field optical response. As a case
study we focus on the interaction of different types of FRBs with
a 100 nm silver spherical NP~\cite{johnson} embedded in glass
(background index $n_\mathrm{b}=1.5$). This is a suitable NP
size since it is neither too small to eliminate the excitation of
higher-order modes, nor too large to suppress the near-field
enhancement~\cite{messinger}. 
There is a freedom of changing $f$ or $w_o$ to obtain different values of 
$a=f/w_o$. We set the beam waist to $w_o=5$ mm and adjust $f$ to select $a$.
We use Mathematica for the calculations~\cite{mathematica}, taking
into account up to 15 multipoles for the field expansions.

\begin{figure}
\begin{center}
\includegraphics[width=8.2cm]{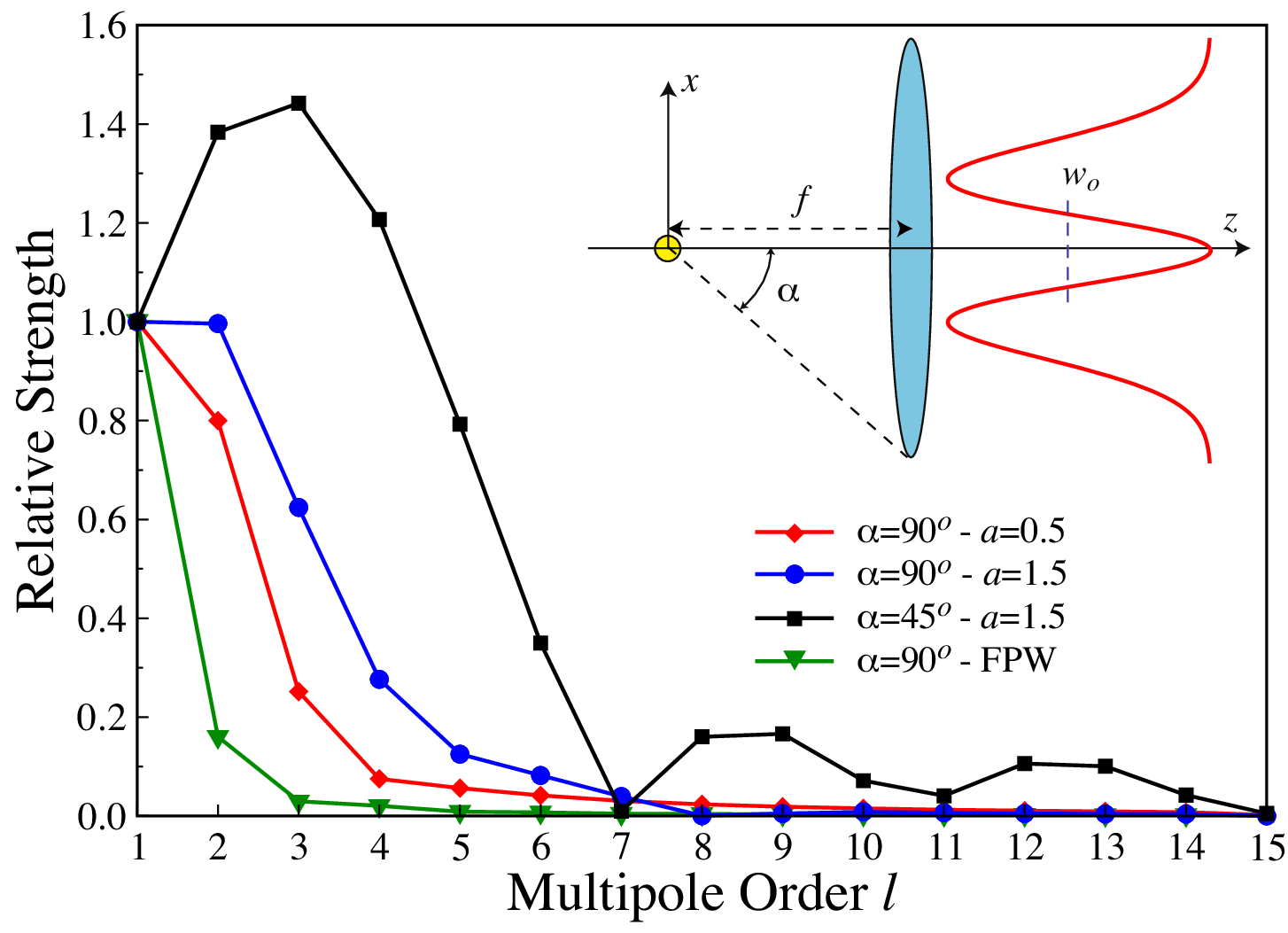}
\caption{Relative strength $|B_l/B_1|$ of the multipoles in the
incident field. The result for a FPW is also shown for 
comparison~\cite{JOSAB}.
The inset sketches a radially-polarized beam focused
onto a NP. The red curve is the beam intensity profile before the 
lens, $f$ is the focal length, $\alpha$ the angular semi-aperture, $w_o$ 
the beam waist and $a=f/w_o$.\label{multipole}} 
\end{center}
\end{figure}

The relative strength $|B_l/B_1|$ in Fig.~\ref{multipole} shows
how much the $l^\mathrm{th}$ multipole contributes to the fields in
comparison to the dipole ($l=1$). A tightly-focused beam
($\alpha=90^o$ and $a=0.5$) mainly consists of a dipole and a
quadrupole, the latter being 80\% of the dipole strength.  The
next case has again the same focusing angle, but the beam waist is
smaller. This is a special illumination because the quadrupole and
the dipole coefficients exhibit nearly the same weight. Keeping
the same value for $a$ but focusing less tightly ($\alpha=45^o$),
the dipole strength becomes even less than the next three multipoles.
In summary, the general trend in the incident-wave multipole
content is as follows: by increasing the value of $\alpha$, i.e. focusing 
more tightly, higher-order modes are suppressed and by increasing the 
value of $a$, i.e. longer focal length or smaller beam waist, 
higher-order modes get stronger. 
For the sake of comparison we also plot in Fig.~\ref{multipole} the 
multipole strength for a FPW. Note that for a FPW and a Gaussian beam the 
higher-order modes can only be suppressed with respect to an incident plane
wave~\cite{JOSAB,lock}. 

\begin{figure}[ht]
\begin{center}
\includegraphics[width=8.2cm]{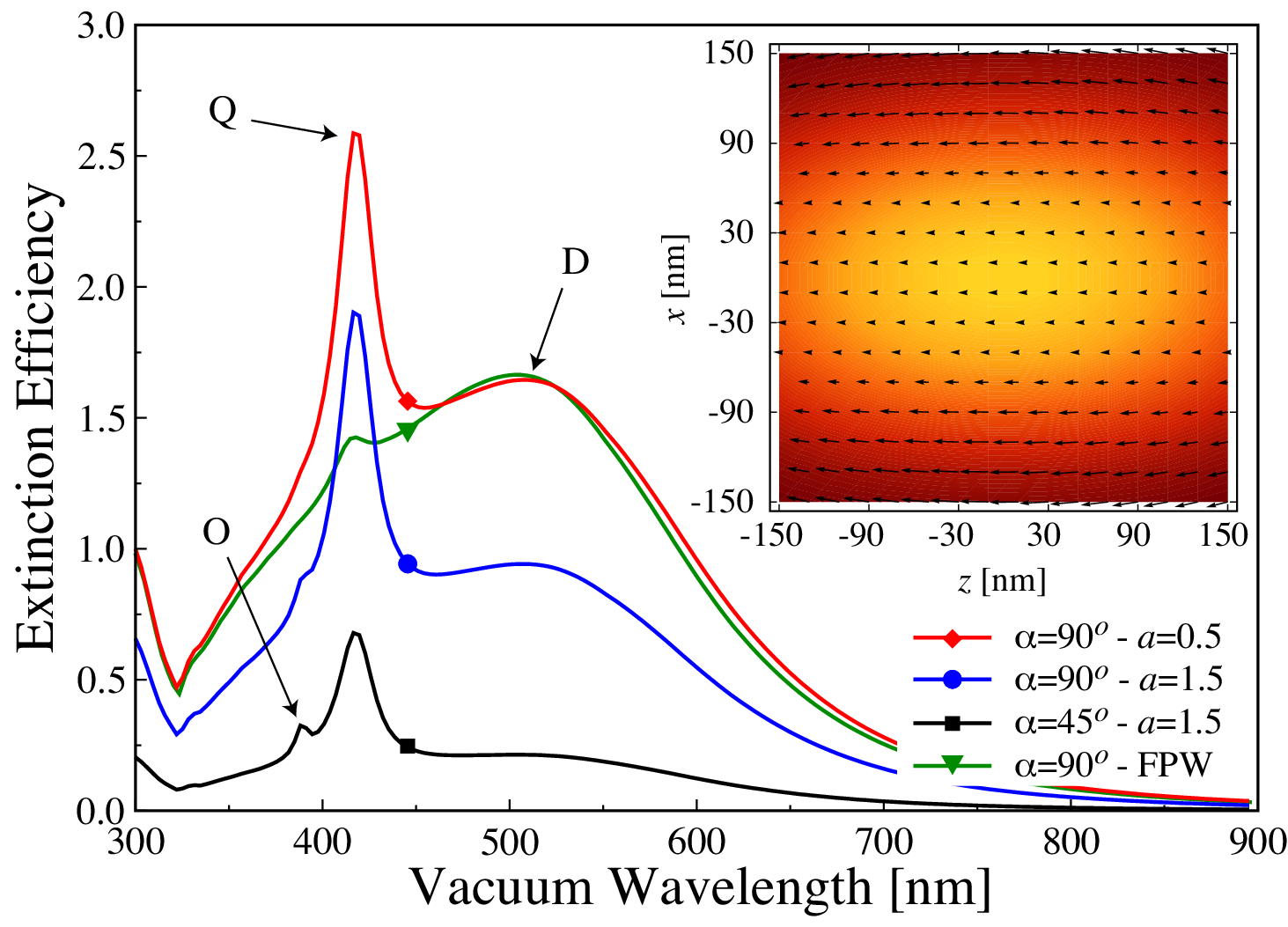}
\caption{Extinction efficiency $\mathcal{K}_\mathrm{e}$ of a 100 nm silver
NP in glass illuminated by three different FRBs and a FPW. D, Q and O 
respectively label the dipole, quadrupole and octupole resonances.
Inset: electric field intensity (contours) and Poynting vector (arrows) 
in the focal region for $\alpha=90^o$, $a=1.5$, and $\lambda=520$ nm. 
\label{ext}}
\end{center}
\end{figure}

We next study the far-field LSP spectrum by looking at 
$\mathcal{K}_\mathrm{e}$ (see Fig.~\ref{ext}). The first property that
catches one's attention is that $\mathcal{K}_\mathrm{e}$ can exceed 2. This
does not violate the conservation of energy, because one has to
consider the total Poynting vector instead of only the scattered or
extinguished components~\cite{bohren,zumofen}. Indeed the transmitted,
reflected, and absorbed powers together are equal to the total input
power. The  effect of changing the multipoles strength can be
observed here by studying the relative heights of the
corresponding resonances. For example, the ratios of the quadrupole
to dipole peaks are 1.63, 2.04, and 3.22 for the three cases
presented in Fig.~\ref{multipole}. As expected, the
multipole strength determines the relative height of the LSP
modes accordingly. Indeed for the FRB with a large content of 
higher-order multipoles, even the third resonance is clearly excited.
To stress the difference with respect to conventional focused light, 
Fig.~\ref{ext} shows that almost only the dipole resonance contributes to 
$\mathcal{K}_\mathrm{e}$ for a FPW when $\alpha=90^o$.

Comparing these three cases reveals that the NP extinguishes more
power when the beam is focused more tightly. This is simply because
more light is concentrated onto the NP. Furthermore, the reason
why $\mathcal{K}_\mathrm{e}$ is larger for a shorter focal length (smaller
$a$) is that the electric field of the rays farther away
from the $z$ axis produce a stronger longitudinal field at the
focus~\cite{leuchs1}. The interesting consequence of the multipole
expansion is the correlation between the excited LSP
resonances and the strength of the corresponding multipoles. This
information can be used to control the excitation of each mode in
the NP.

\begin{figure}
\begin{center}
\includegraphics[width=8.2cm]{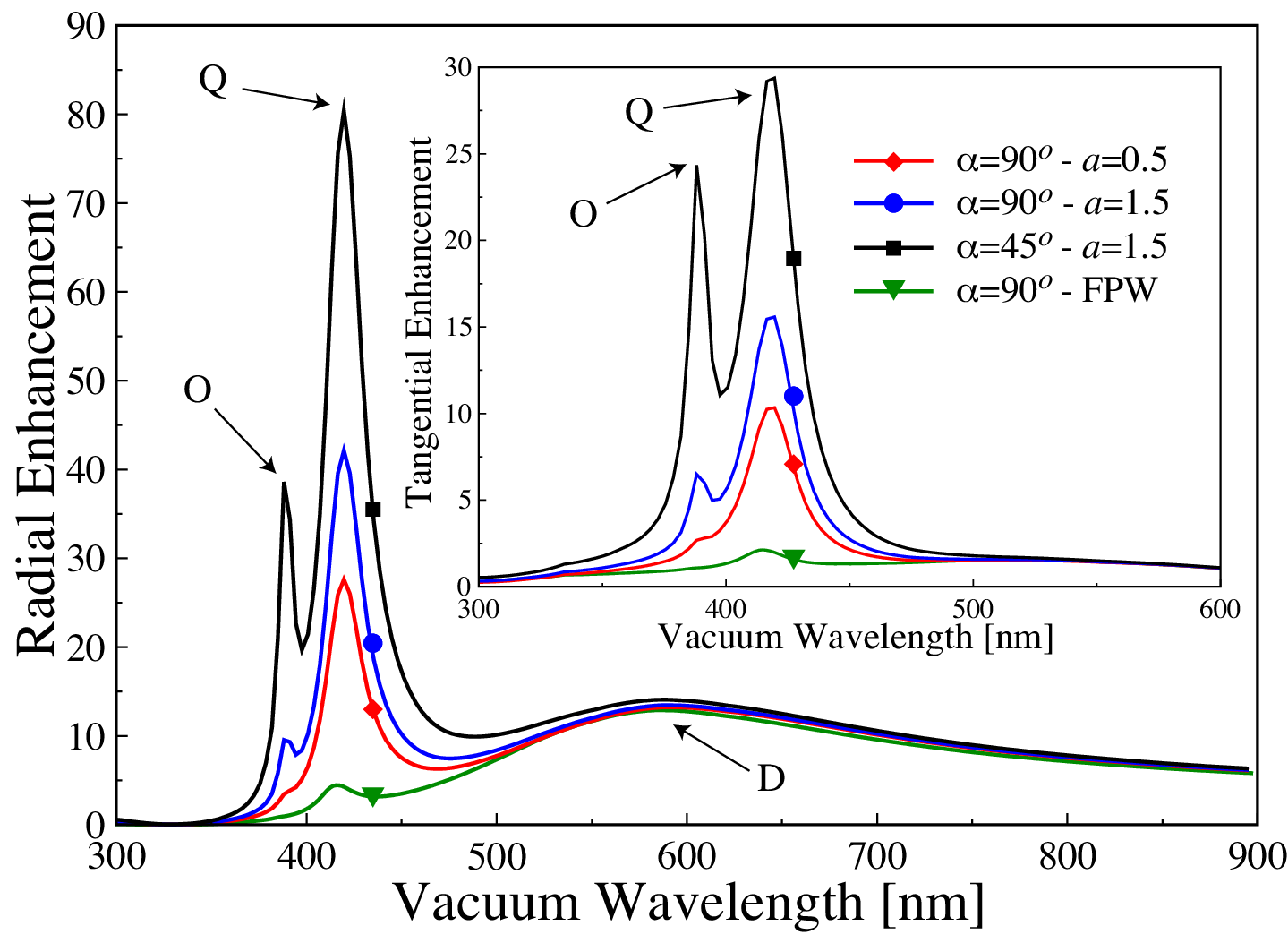}
\caption{Radial and tangential (inset) average intensity
enhancement for a 100 nm silver NP in glass illuminated by three
different FRBs and a FPW. D, Q and O respectively label the dipole, 
quadrupole and octupole resonances.\label{int}}
\end{center}
\end{figure}

We next move our attention to the average near-field intensity
enhancement and especially on the radial one (see Fig.~\ref{int}).
For a 100 nm silver NP, the higher-order excited modes yield
better $K_\mathrm{r}$ and $K_\mathrm{t}$ with respect to the dipole one.
Moreover, by focusing less tightly or making $a$ larger, they exhibit
a stronger enhancement.
In comparison to $\mathcal{K}_\mathrm{e}$ of Fig.~\ref{ext},
the relative strength of the quadrupole peak with
respect to the dipole is larger. This clearly shows the difference
between the near- and the far-field regimes~\cite{JOSAB,messinger}.
As in the case of FPW illumination, which is shown in Fig.~\ref{int}
for comparison, the enhancement near the dipole peak remains almost the same
irrespective of the focusing parameters. 
For a FPW $K_\mathrm{r}$ and $K_\mathrm{t}$ are always less than for 
plane-wave illumination~\cite{JOSAB}. 
On the other hand, with FRBs one can control the enhancement at the 
quadrupole and higher-order resonances by changing the multipole content 
of the beam. More precisely,
the ratio of quadrupole to dipole peak enhancements is 2.08, 3.13,
and 5.71 for the three cases in Fig.~\ref{int}. For
instance, when $\alpha=45^o$ and $a=1.5$ $K_\mathrm{r}$
reaches a factor of 80, which is about 2 and 6 times the value 
for a plane wave and a FPW for $\alpha=60^o$, respectively~\cite{JOSAB}.

\section{Conclusion}
We have studied the near- and far-field response of metal NPs
illuminated by FRBs using a 100 nm silver spherical NP in glass as
a representative system. The multipole expansion approach allows
us to demonstrate the potential of FRBs for tailoring the
excitation of LSP resonances and hence controlling
the scattering and extinction efficiencies or the near-field
enhancement. These findings show that FRBs offer more
opportunities than conventional focused light.
Our results might find application in the
spectroscopy of metal NPs~\cite{meixner3}, optical
tweezers~\cite{zhan1} and optical data storage~\cite{sugiyama}.
Furthermore, molecular fluorescence can also be improved by FRBs.
Since the quadrupole mode exhibits a stronger electric field
under appropriate illumination, one could use it to
increase the excitation rate and the dipole resonance to amplify the
emission rate without the need of designing complicated
nanostructures~\cite{giannini}.

\section*{Acknowledgments}
We thank V. Sandoghdar for continuous support and
encouragement. We are also indebted to G. Zumofen and M.~H.
Eghlidi for fruitful discussions. This work was supported by ETH
Zurich.


\end{document}